\RequirePackage{ifpdf}
\ifpdf 
\documentclass[pdftex]{sigma}
\else
\documentclass{sigma}
\fi

\def\d{\mbox{\rm d}}

\def\half{\mbox{$\frac{1}{2}$}}

\def\dddot#1{\mbox{\small$\mathinner{\buildrel\vbox{\kern4pt\hbox{...}}\over{#1}}$}}

\begin{document}

\renewcommand{\PaperNumber}{018}

\FirstPageHeading

\ShortArticleName{Ermakov's Superintegrable Toy and Nonlocal
Symmetries}

\ArticleName{Ermakov's Superintegrable Toy \\ and Nonlocal
Symmetries}

\Author{P.G.L. LEACH~${}^{\dag^1}$, A. KARASU (KALKANLI)~${}^{\dag^2}$, 
M.C. NUCCI~${}^{\dag^3}$ and K. ANDRIOPOULOS~${}^{\dag^4}$}
\AuthorNameForHeading{P.G.L.~Leach, A.~Karasu (Kalkanl\i), M.C.~Nucci and K.~Andriopoulos}

\Address{${}^{\dag^1}$~School of Mathematical Sciences, Howard College, 
University of KwaZulu-Natal,\\
$\phantom{{}^{\dag^1}}$~Durban 4041, Republic of South Africa} 
\EmailDD{\href{mailto:leachp@ukzn.ac.za}{leachp@ukzn.ac.za}}

\Address{${}^{\dag^2}$~Department of Physics, Middle East Technical
University, 06531 Ankara, Turkey}
\EmailDD{\href{mailto:akarasu@metu.edu.tr}{akarasu@metu.edu.tr}}

\Address{${}^{\dag^3}$~Dipartimento di Mathematica e Informatica,
Universit\`a di Perugia, 06123 Perugia, Italy}
\EmailDD{\href{mailto:nucci@unipg.it}{nucci@unipg.it}}

\Address{${}^{\dag^4}$~Department of Information
and Communication Systems Engineering,\\
$\phantom{{}^{\dag^1}}$~University of the Aegean, Karlovassi 83 200, Greece}
\EmailDD{\href{mailto:kand@aegean.gr}{kand@aegean.gr}}

\ArticleDates{Received September 19, 2005, in final form November 11, 2005;
Published online November 15, 2005}

\Abstract{We investigate the symmetry properties 
of a pair of Ermakov equations.  The system is superintegrable 
and yet possesses only three Lie point symmetries with the algebra $sl(2,{\mathbb R}) $.  
The number of point symmetries is insufficient and the algebra unsuitable 
for the complete specification of the system.  
We use the method of reduction of order to reduce the
nonlinear fourth-order system to a third-order system comprising 
a linear second-order equation and a conservation law.  
We obtain the representation of the complete symmetry group from this system. 
Four of the required symmetries are nonlocal and the algebra is the direct 
sum of a one-dimensional Abelian algebra with the semidirect sum of a 
two-dimensional sol\-vable algebra with a two-dimensional Abelian algebra.  
The problem illustrates the difficulties which can arise in very elementary systems.  
Our treatment demonstrates the existence of possible routes 
to overcome these problems in a systematic fashion.}


\Classification{17B80; 22E70; 34C14; 37C80}

\section{Introduction}

In 1880 V.~Ermakov \cite{Ermakov 80 a} 
introduced a system of second-order ordinary differential equations, {\it videlicet}
\begin{subequations}\label{1.1/1.2}
\begin{gather}
\ddot {x} +\omega ^ 2 (t)x  =  \displaystyle {\frac {1} {x ^ 3}}, \label {1.1}\\
\ddot {y} +\omega ^ 2 (t)y  =  0, \label {1.2}
\end{gather}
\end{subequations}
for which an invariant could be constructed by means of the
operations $x \cdot (\ref {1.2}) - y \cdot (\ref {1.1})$, multiplication by
an integrating factor $x \dot {y} -\dot {x}y $, and integration
with respect to time.  The invariant obtained is
\begin{gather}
I = \half\left [\left (x\dot {y} -\dot {x}y\right) ^ 2
+\left (\frac {y} {x}\right) ^ 2\right]. \label {1.3}
\end{gather}
In the context of Ermakov's system the dependent variables $x $ and $y $ 
could be interpreted as spatial variables and the independent variable $t $
 as time although this is not necessary.  In such an interpretation the Ermakov 
 system is a pair of coupled oscillators, one of them being nonlinear, with time-dependent 
 frequency~$\omega ^ 2 (t) $.  The Ermakov invariant is then a generalisation 
 of the conservation of angular momentum~\cite{Leach 91 a}.

In another context and era the motion of a charged particle in an 
axially symmetric electromagnetic field was modelled in the linear 
approximation as the time-dependent linear oscillator
\begin{gather}
\ddot {q} +\omega ^ 2 (t)q = 0. \label {1.4}
\end{gather}
By an application of Kruskal's asymptotic method \cite {Kruskal 62
a} Lewis \cite {Lewis 67 a, Lewis 68 a, Lewis 68 b} obtained an
invariant
\begin{gather}
I = \half\left [\left (\rho p-\dot {\rho}q\right) ^ 2+\left(\frac {q} {\rho}\right) ^ 2\right] \label {1.5}
\end{gather}
in which $\rho (t) $ is a solution of the auxiliary equation
\begin{gather}
\ddot {\rho} +\omega ^ 2 (t)\rho = \frac {1} {\rho ^ 3} \label {1.6}
\end{gather}
and $p $ is the momentum canonically 
conjugate to $q $.  Lewis and Riesenfeld \cite{Lewis 69 a} 
provided a~quantum mechanical treatment of the time-dependent linear
 oscillator through the invariant~(\ref {1.5}) as a~linear operator 
 in the context of the Schr\"odinger Equation.  This implied connection of~$I $ 
 with an~Hamiltonian operator was confirmed by Leach~\cite{Leach 77 a} 
 who related the Hamiltonian of the time-dependent linear oscillator
\begin{gather*}
H = \half\left [p ^ 2+\omega ^ 2 (t)q ^ 2\right] 
\end{gather*}
with $I $ through a time-dependent linear generalised canonical
transformation \cite{Burgan 78 a, Burgan 78 b}
\begin{gather*}
Q = \frac {q} {\rho},\qquad P = \rho p-\dot {\rho}q,\qquad T = \int\rho ^ {- 2} (t)\d t, 
\end{gather*}
where $\rho (t) $ is again given by~(\ref {1.6}).

The approaches of Ermakov and Lewis are different even though they
embrace the same three mathematical objects.  The order of
introduction is different.  Ermakov has a system, which could in
principle be coupled although this is not found in the original
treatment, and the system leads to an invariant.  Lewis has a
single equation and an invariant which relies upon the existence
of a second equation. The modes of interpretation, the Ermakov
invariant as a generalisation of the conservation of the magnitude
of angular momentum and the Lewis invariant as a type of
energylike Hamiltonian, reflect the differences in
approach\footnote {In this respect the interpretation by Eliezer
and Gray \cite {Eliezer 76 a} of the Lewis invariant as a
generalisation of angular momentum was rather prescient in that it
preceeded by a few years the spread of knowledge of Ermakov
systems beyond the region of Kiev.  That G\"unther {\it et al}
\cite {Gunther 77 a} could demonstrate the failure of the
interpretation when the number of spatial dimensions was increased
underlies the fortuitous nature of the interpretation.}.

A further complication had already been provided by Pinney \cite{Pinney 50 a, Notices 02 a}
 who related the solutions of the auxiliary equation (\ref{1.6})
 to that of the time-dependent linear oscillator (\ref{1.4}) according to
\begin{gather}
\rho(t) = \left(Au^2+2Buv+Cv^2\right)^{\half}, \label{1.9}
\end{gather}
where $u(t)$ and $v(t)$ are any two linearly independent solutions
of (\ref{1.4}) and the three constants $A$, $B$ and $C$ are
related according to $AC-B^2 = 1/W^2$, where $W:=
u\dot{v}-\dot{u}v$ is the constant Wronskian of the basis of
solutions of (\ref{1.4}).

More recent literature \cite {Mahomed 90 a, Moyo 02 a, Andriopoulos
05 b} links the solution (\ref {1.9}) to that of the linear
third-order differential equation
\begin{gather}
\dddot {y} +4\omega ^ 2 (t)\dot {y} +4\omega (t)\dot {\omega} (t)y = 0, \label {1.10}
\end{gather}
where $y = \half\rho ^ 2 $, and its integration by means of the
integrating factor~$y $.  This connection becomes even more
manifest when one recognises that (\ref {1.10}) is the equation
satisfied by the coefficient functions of the three generators of
the algebra $sl (2,{\mathbb R}) $ possessed by all linear $n $th-order
ordinary differential equations of maximal symmetry\footnote {There
are some variations in the precise forms of the coefficients of
$\dot {y} $ and $y $ to allow for the value of $n $; see Mahomed
{\it et al} \cite{Mahomed 90 a}.  Indeed the third-order equation~(\ref{1.10})
 as adjusted for the order of the equation under
study, provides a unifying feature to the study of the Lie algebras
of $n $th-order ordinary differential equations as it is always
present for equations of maximal symmetry.  For $n\geq 3 $ the other
symmetries come from the solution of an $n $th-order differential
equation and a first-order differential equation.  The solutions of
the $n $th-order equation and third-order equation are related to
the solutions of a second-order equation of maximal symmetry 
\cite{Moyo 00 a}.  In fact this relationship leads to a simple
demonstration of Pinney's solution (\ref {1.9}).  For $n = 2 $
another second-order differential equation intrudes to provide the
noncartan symmetries.  This second-order differential equation is
the adjoint of the second-order equation under study 
\cite{Turkmen 05 a}.}  and that $sl (2,{\mathbb R}) $ is also the algebra of the Lie point
symmetries of the Ermakov--Pinney equation (\ref{1.1}), (\ref{1.6}).

Around 1978 the systems of Ermakov became known in the
 West -- there is more than a little folklore about the mode of dissemination -- 
 and immediately attracted a great deal of attention; for a modest sampling see 
 \cite{Ray 80 a, Ray 80 b, Ray 80 c}.  The flood of papers did diminish somewhat,
  but more recent articles \cite {Haas a, Haas b, Haas c} 
  indicate that the mathematical and physical riches of Ermakov systems have not been exhausted.

In this paper we wish to treat a two-dimensional Ermakov system in a
manner analogous to the treatment of the `beloved' equation\footnote
{`L'\'equation ador\'e' as (\ref {1.11}) was dubbed, perhaps a
little irreverently, by the recently and sadly late Mark Feix in 1996 after yet another
attractive property of this remarkable equation was revealed in the
work of his student, Claude G\'eronimi.}
\begin{gather}
\dddot {x} + 3x\dot {x} + x ^ 3 = 0 \label {1.11}
\end{gather}
by Karasu {\it et al} \cite{Karasu 05 a} in which a doubled
version of (\ref {1.11}) was studied from the viewpoint of its
symmetries and integrability in the sense of Painlev\'e.  The
collapse of the Lie point symmetry algebra from the single (\ref{1.11}) 
to its dual was dramatic.  In that treatment there existed
a linea\-rising transformation to a pair of third-order ordinary
differential equations and the Lie point symmetries of the
third-order equations enabled the complete symmetry group of the
system of second-order equations to be obtained in terms of a set
of five nonlocal symmetries which was distinguished by being
Abelian.

The complete symmetry group of a system of ordinary differential
equations is the group of the algebra of Lie symmetries required
to specify the system completely.  The concept was introduced by
Krause \cite{Krause 94 a, Krause 95 a} in a study of the Kepler
Problem.  The absence of a sufficient number of Lie point
symmetries for the purpose caused Krause to introduce nonlocal
symmetries by an ingenious strategy which has proved to be useful
elsewhere \cite{Nucci 02 b}.  The necessity for the introduction
of an Ansatz for the structure of nonlocal symmetries was
successfully challenged by Nucci \cite {Nucci 96 a} who used the
method of reduction of order \cite {Nucci 00 a} to obtain the
requisite number of symmetries wholly through point symmetries --
naturally with a coupling to a nonlocal transformation in the case
of those symmetries which were nonlocal in the original
representation. This same method has been found to apply to a wide
variety of systems \cite{Leach 02 b, Nucci 04 a} and has revealed
an unanticipated algebraic identity of somewhat dissimilar
problems. The present theoretical development of the properties of
complete symmetry groups can be found in \cite{Andriopoulos 01 a,
Andriopoulos 02 a, Andriopoulos 02 b, Andriopoulos 05 a}.

The system which we treat is
\begin{subequations}\label{1.12/1.13}
\begin{gather}
\ddot {x}  =  \displaystyle {\frac {1} {x ^ 3}}, \label {1.12}\\
\ddot {y}  =  \displaystyle {\frac {1} {y ^ 3}} \label {1.13}
\end{gather}
\end{subequations}
in which each component is the simplest form of the Ermakov--Pinney
equation possible and no coupling between the two equations
exists.  The reader sceptical of the potential difference between
a single Ermakov--Pinney equation and a pair cloned as above is
referred to the system discussed by Karasu {\it et al}~\cite{Karasu 05 a} in which these
differences are displayed in detail.  We keep the system at its
simplest to avoid masking the clarity of the concept under
consideration behind a~complexity of model\footnote{In a sense
this already happened with complete symmetry groups which entailed
the use of nonlocal symmetries \cite{Krause 94 a, Krause 95 a}
whereas the idea can be displayed in the context of point
symmetries and the most elementary of ordinary differential
equations \cite{Andriopoulos 01 a, Andriopoulos 02 a,
Andriopoulos 02 b}.}.  We note that (\ref{1.12/1.13}) is
related to the more usual form of Ermakov systems by means of a
point transformation \cite{Leach 91 a}.

We examine the system (\ref{1.12/1.13}) from the point of
view of its Lie point symmetries and find that these are
insufficient to specify completely the system.  A recent work of
Nucci {\it et al}~\cite{Nucci 05 b} on the complete symmetry group
of (\ref{1.12}) does provide an easy route to the complete symmetry
group of the system (\ref{1.12/1.13}) although the ease does
follow from the prior passage of the route.  For a~single
Ermakov--Pinney equation that route required the application of the
method of reduction of order.  The method relies upon the existence
of an ignorable coordinate.  In the case of 
(\ref {1.12/1.13}) time is the obvious ignorable coordinate.  The choice of a
new independent coordinate is not obvious\footnote{In the case of
just (\ref{1.12}) this is obviously~$x $.}.  However, a transition
to polar coordinates makes the choice traditionally obvious.  In
polar coordinates the method of reduction of order leads to the
combination of a linear second-order ordinary differential equation
and a trivial first-order differential equation, i.e.\ a~conservation 
law, reminiscent of the reduction of the MIC-Z Problem
\cite{Nucci 04 b} in addition to problems more obviously connected
to the Kepler Problem.

The reduction provides a source of Lie point symmetries which can
then be translated to the original coordinates and equations to
give a suite of symmetries from which the complete symmetry group
of system (\ref{1.12/1.13}) can be determined. Fortunately
the theorems developed by Andriopoulos {\it et al}~\cite{Andriopoulos 05 a} 
obviate the necessity for lengthy calculations
involving nonlocal symmetries for we can use the symmetries of the
complete symmetry group of the derivate system and the transitive
properties of complete symmetry groups to arrive at the complete
symmetry group of (\ref{1.12/1.13}).  We note that in this
derivation the nonuniqueness of the representation given by Nucci
{\it et al} is replicated due to the known property of the
possession by linear second-order ordinary differential equations
of at least three representations of their complete symmetry groups 
\cite{Leach 88 a, Andriopoulos 02 a}.

The interesting feature of system (\ref{1.12/1.13}) 
is that it is rather light on Lie point symmetries and yet, as we see below, 
is a superintegrable system in the standard sense in that it possesses three first integrals.

The essential points of this paper fall into two parts.  Individually (\ref {1.12}) 
and (\ref {1.13}) are endowed with sufficient symmetry to ensure an easy integration.  
In this context we mean sufficient Lie point symmetries as is the standard for second-order 
equations and usual for higher order equations.  The simple cojoining 
of the two equations produces the situation in which the number of 
Lie point symmetries is formally fewer than that required for easy 
integration and yet the ease of integration remains.  We use this simple system 
to emphasise the problems of moving from a~scalar equation to a system of equations.  
The second part of the paper deals with the solution of the problem posed by the 
lack of Lie point symmetries by comparison with the individual scalar second-order equations.  
The resolution of such problems is not necessarily by means of a~recipe.  Were this so, 
the solution of systems of nonlinear equations would be no more difficult 
than that of scalar equations.  Anyone with the slightest acquaintance 
with the problem of the solution of systems of nonlinear equations knows 
the foolishness of that position.  What we do in this paper reflects the two parts.  
Firstly we delineate the problems encountered in the simple act of cojoining 
two integrable equations as far as the usual criteria of integrability in the 
Lie theory are concerned.  Secondly we demonstrate how those problems are overcome.  
The particular resolution of these problems may be unique to the system, 
but the techniques are of a general nature and belong to the  
standard repertoire to be used when a system displays refractory qualities.  
The techniques used here were not designed for this system.  
They exist and may be employed generally.  The success of their 
employment depends upon the specific nature of the system under consideration.

We emphasise that the system we study (\ref{1.12/1.13}) 
is quite artificial and is studied more to illustrate properties 
and methods than to provide novel application.  
Yet we find it an appropriate object for our attentions in 
this 125th anniversary of its genesis in this noble city of Kiev by one of its now well-known sons.

\section{The superintegrable toy}

There are three Lie point symmetries of system (\ref{1.12/1.13}), {\it videlicet}
\begin{gather}
\Gamma_1   =   \partial_t, \nonumber\\
\Gamma_2   =   t\partial_t+\half (x\partial_x+y\partial_y), \label{2.1}\\
\Gamma_3  =  t ^ 2\partial_t+ t (x\partial_x+y\partial_y),
\nonumber
\end{gather}
with the Lie Brackets
\begin{gather*}
\left [\Gamma_1,\Gamma_2\right] = \Gamma_1,\qquad \left [\Gamma_1,\Gamma_3\right]
 = 2\Gamma_2,\qquad \left [\Gamma_2,\Gamma_3\right] = \Gamma_3 
\end{gather*}
characteristic of the algebra $sl(2,{\mathbb R}) $.  The system (\ref{1.12/1.13})
 may be regarded as a pair of first integrals
consequent upon the use of the integrating factors $\rho $ and
$\sigma $, respectively, of the system
\begin{gather}
\begin{array}{l}
\dddot {\rho}  =  0,\\
\dddot {\sigma}  =  0,
\end{array}\Longrightarrow
\begin{array}{l}
\rho\dddot {\rho} = 0,\\
\sigma\dddot {\sigma} = 0,
\end{array}\Longrightarrow
\begin{array}{l}
\displaystyle \rho\ddot {\rho} -\half \dot{\rho} ^ 2 = \half A,\vspace{2mm}\\
\displaystyle \sigma\ddot {\sigma} -\half \dot{\sigma} ^ 2 = \half B,
\end{array}\Longrightarrow
\begin{array}{l}
\rho = \half x ^ 2,\vspace{2mm}\\
\sigma = \half y ^ 2,
\end{array}\Longrightarrow
\begin{array}{l}
x ^ 3\ddot {x} = A,\\
y ^ 3\ddot {y} = B
\end{array}
\label {2.3}
\end{gather}
with the specific choice of the constants of integration $A $ and $B $ being unity.  
The third-order system is of maximal symmetry with
\begin{gather}
\begin{array} {lcl}
\Lambda_{11} = \partial_{\rho},& & \Lambda_{21} = \partial_{\sigma},\vspace{1mm}\\
\Lambda_{12} = t\partial_{\rho}, & &\Lambda_{22} = t\partial_{\sigma}, \vspace{1mm}\\
\Lambda_{13} = \half t ^ 2\partial_{\rho}, & &\Lambda_{23} = \half t ^ 2\partial_{\sigma}, \vspace{1mm}\\
\Lambda_{14} = \rho\partial_{\rho}, & &\Lambda_{24} = \sigma\partial_{\sigma}, \vspace{1mm}\\
\Lambda_{15} = \sigma\partial_{\rho},& &\Lambda_{25} = \rho\partial_{\sigma}, \vspace{1mm}\\
&\Lambda_{c1} = \partial_t, &\vspace{1mm}\\
& \Lambda_{c2} = t\partial_t, &\vspace{1mm}\\
& \Lambda_{c3} = t ^ 2\partial_t+ 2t (\rho\partial_{\rho} +\sigma\partial_{\sigma}) &
\end{array} \label {2.4}
\end{gather}
and only three of these survive in the integrals whether 
they be treated as functions or equations\footnote{System (\ref {2.3}) 
has the usual ambiguity of self-similar symmetries for linear systems of maximal symmetry. 
 The nonlinearity of (\ref{1.12/1.13})
enforces the precise form of $\Gamma_2 $ in~(\ref {2.1}).  
In terms of the variables $\rho $ and $\sigma $ the 
symmetries $\Gamma_2 $ and $\Gamma_3 $ are in the standard form for 
the representation of $sl (2,{\mathbb R}) $ appropriate to equations of the third order.  
The transformation to $x $ and $y $ gives the standard form for second-order equations 
\cite{Mahomed 90 a}, but we do well to recall that the 
symmetries are from the third-order system (\ref{2.3}) as listed in (\ref{2.4}).}.

Two first integrals of system (\ref{1.12/1.13}) are obviously
\begin{gather*}
I  =  \half\left (\dot {x} ^ 2+\displaystyle {\frac {1} {x ^ 2}}\right) \qquad
\mbox{\rm and}\qquad
J  =  \half\left (\dot {y} ^ 2+\displaystyle {\frac {1} {y ^
2}}\right). 
\end{gather*}
The third first integral is the Ermakov invariant
\begin{gather}
K = \half\left [\left (x\dot {y} -\dot {x}y\right) ^ 2
+\frac {x ^ 2} {y ^ 2} +\frac {y ^ 2} {x ^ 2}\right] \label {2.6}
\end{gather}
which is constructed in the same manner as for the Ermakov 
invariant (\ref {1.3}) of system~(\ref{1.1/1.2}).  System~(\ref{1.12/1.13}) 
is of the fourth order and has three autonomous constants of the motion.  
Hence system~(\ref{1.12/1.13}) is superintegrable.

The autonomous integrals can all be derived from the 
symmetry $\partial_t$ if one uses Lie's method (cf.~\cite{Leach 80 a}).  
Invariance under $\Gamma_1$ means that the integral is autonomous, i.e.\ 
of the form $F(x,y,\dot{x},\dot{y})$. 
 The associated Lagrange's system for $\d F/\d t = 0$ is
\begin{gather}
\frac{\d x}{\dot{x}} = \frac{\d y}{\dot{y}} = \frac{\d
\dot{x}}{x^{-3}} = \frac{\d \dot{y}}{y^{-3}} \label{2.7}
\end{gather}
when (\ref{1.12/1.13}) are taken into account.  
The combination of the first and third of (\ref{2.7}) 
leads to~$I$ and of the second and fourth to~$J$.  The determination 
of $K$ is a little more complicated.  The combinations 
$\dot{y}\cdot(\ref{2.7}{\rm a}) - \dot{x}\cdot (\ref{2.7}{\rm c}) + 
y\cdot (\ref{2.7}{\rm b})-x\cdot (\ref{2.7}{\rm d})$ and $y\cdot (\ref{2.7}{\rm a})-x\cdot(\ref{2.7}{\rm b})$ give
\begin{gather*}
\frac{\d\left(x\dot{y}-\dot{x}y\right)}{\dfrac{x}{y^3}-
\dfrac{y}{x^3}} = \frac{y\d x-x\d y}{y\dot{x}-x\dot{y}} 
\end{gather*}
from which the third integral $K$ follows.

Although $\Gamma_1$ generates the three autonomous first integrals
of system (\ref{1.12/1.13}), each integral possesses an
infinite number of Lie point symmetries.  In the case of $K$ these
are \cite{Govinder 93 b, Govinder 94 c}
\begin{gather*}
\Sigma_K = f(t)\partial_t+\half\dot{f}(t)\left(x\partial_x+y\partial_y\right), 
\end{gather*}
where $f(t)$ is an arbitrary, differentiable, function.  
In the cases of $I$ and $J$ the infinite classes of Lie point symmetries are
respectively
\begin{gather*}
\Sigma_I = \partial_t + f(t,x,y)\partial_y, \\ 
\Sigma_J = \partial_t + f(t,x,y)\partial_x 
\end{gather*}
which are both perhaps to be expected and not as interesting as in the case of~$K$.

As an amusing endpoint to this discussion of the integrals we note that 
the integrals associated with $\Gamma_3$ are
\begin{gather*}
I_3 = \half\left\{\left(t\dot{x}-x\right)^2+\left(\frac{t}{x}\right)^2\right\}, \qquad 
J_3 =
\half\left\{\left(t\dot{y}-y\right)^2+\left(\frac{t}{y}\right)^2\right\}
\end{gather*}
and $K$ as given in (\ref{2.6}).  If one uses the point transformation
\begin{gather}
T = -\frac{1}{t}, \qquad  X =\frac{x}{t}, \qquad Y = \frac{y}{t}, \label{2.14}
\end{gather}
the integrals $I$ and $J$ are recovered.  (Obviously $K$ is
invariant under the point transformation.)  The transformation
(\ref{2.14}) is a simple instance of the generalised canonical
transformation \cite{Burgan 78 a, Burgan 78 b}
\[
T = \int \rho^{-2}(t)\d t, \qquad Q = \frac{q}{\rho}, \qquad P =\rho
p-\dot{\rho}q
\]
which has found much application in both Hamiltonian and Quantum
Mechanics \cite{Leach 77 a, Lewis 82 a, Pillay 98
a}\footnote{Apart from $K$ the symmetry $\Gamma_2$ has the two
integrals $I_2 = \dot{x}(t\dot{x}-x) + t/x^2$ and $J_2 =
\dot{y}(t\dot{y}-y) + t/y^2$.  This in itself is interesting as
the three integrals $I_1$, $I_2$ and $I_3$, respectively $J_1$,
$J_2$ and $J_3$, are generalisations of the quadratic integrals of
the one-dimensional free particle which for $\ddot{x}=0$ are
$\half\dot{x}^2$, $\dot{x}(t\dot{x}-x)$ and $\half(t\dot{x}-x)^2$.
These integrals are obviously products of the two linear
integrals $\dot{x}$ and $t\dot{x}-x$.  The nonlinear Ermakov
system does not possess linear integrals.  However, the quadratic
integrals could be considered as the inner products of the
nonconserved vectors ${\boldsymbol i}_x = \dot{x}\hat{{\boldsymbol a}}_1 +
\frac{1}{x}\hat{\boldsymbol a}_2$, ${\boldsymbol i}_y = \dot{y}\hat{{\boldsymbol e}}_1 +
\frac{1}{y}\hat{\boldsymbol e}_2$, ${\boldsymbol j}_x = (t\dot{x}-x)\hat{{\boldsymbol
a}}_1 + \frac{t}{x}\hat{\boldsymbol a}_2$ and ${\boldsymbol j}_y =
(t\dot{y}-y)\hat{{\boldsymbol e}}_1 + \frac{t}{y}\hat{\boldsymbol e}_2$, 
where~$\hat{\boldsymbol a}_i$ and $\hat{\boldsymbol e}_i$, $i=1,2$, are unit vectors. In
further contrast to the free particle the inner products of the
$x$ vectors and the $y$ vectors are not conserved.}.

\section {Complete symmetry group}

The Lie algebra $sl (2,{\mathbb R}) $ 
is not found as a constituent of a complete symmetry group since the 
Lie Brackets show a redundancy of operators and this thereby 
leads to a violation of the condition of minimality which is one of 
the requirements for a representation of a complete symmetry group \cite{Andriopoulos 01 a, Nucci 05 b}.  
This then poses the problem of finding the elements of the algebra of the complete symmetry 
group for (\ref{1.12/1.13}) since the only Lie point symmetries are the elements of $sl (2,{\mathbb R}) $.  
Karasu {\it et al}~\cite{Karasu 05 a}, when
 confronted with a similar obstacle to progress in determination, 
 made use of a nonlocal transformation to a third-order system of the form of (\ref{2.3}) 
 and were able to use the nonlocal forms of the remainder of (\ref{2.4}) 
 at the second-order level to express the complete symmetry group of the system 
 in terms of a five-dimensional Abelian algebra of nonlocal symmetries.

That option is not so apparent for the present system.  However,
we may make use of an extension of the work of Nucci {\it et al}
\cite{Nucci 05 b} to obtain indirectly a representation of the
complete symmetry group for the system (\ref{1.12/1.13}).
They, by a complex sequence of analyses and transformations
facilitated by some recent theorems of Andriopoulos {\it et al}
\cite{Andriopoulos 05 a}, were able to represent the complete
symmetry group of a single Ermakov--Pinney equation.  In the
notation adopted here, (\ref{1.12}) say, is completely described
by the three symmetries
\begin{gather}
\Delta_1  =  \partial_t, \qquad
\Delta_2  =  \left (\int \frac {\d t} {\dot {x} ^ 2}\right)\partial_t, \qquad
\Delta_3  =  \frac 12 \left (\int\left (1+\frac {1} {x
^ 2\dot {x} ^ 2}\right)\d t\right)\partial_t. \label{3.1}
\end{gather}
It is a simple matter to infer that the system (\ref{1.12/1.13}) is completely specified by (\ref{3.1}) plus
\begin{gather}
\Delta_4  =  \left (\int\frac {\d t} {\dot {y} ^ 2}\right)\partial_t, \qquad 
\Delta_5  =  \frac 12 \left (\int\left (1+\displaystyle {\frac {1} {y
^ 2\dot {y} ^ 2}}\right)\d t\right)\partial_t  \label {3.2}
\end{gather}
to give the five symmetries expected by the general theory \cite{Andriopoulos 02 a}.

We verify that $\Delta_1-\Delta_5 $ are a valid representation of the complete symmetry group by considering the actions of $\Delta_1 ^ {[2]}-\Delta_5 ^ {[2]} $ on the general two-dimensional system
\begin{gather*}
\ddot {x}  =  f\left (t,x,y,\dot {x},\dot {y}\right), \nonumber\\
\ddot {y}  =  g\left (t,x,y,\dot {x},\dot {y}\right). 
\end{gather*}

The action of $\Delta_1$ removes $t$ from both of $f$ and $g$.
The second extensions of $\Delta_2$ and $\Delta_4$ are
\[
\Delta_2^{[2]} = \left(\int\frac{1}{\dot x^2}\d t\right)\partial_t -
\frac{1}{\dot{x}}\partial_{\dot{x}}\qquad\mbox{\rm and }\qquad
\Delta_4^{[2]} = \left(\int\frac{1}{\dot y^2}\d t\right)\partial_t -
\frac{1}{\dot{y}}\partial_{\dot{y}}.
\]
They remove $\dot{x}$ and $\dot{y}$ from $f$ and $g$.  The relevant
parts of the second extensions of $\Delta_3$ and $\Delta_5$ are
\[
\Delta_3^{[2]} =
\left\{-f+\frac{1}{x^3}\right\}\partial_{\ddot{x}}\qquad\mbox{\rm
and}\qquad \Delta_5^{[2]} =
\left\{-g+\frac{1}{y^3}\right\}\partial_{\ddot{y}}
\]
from which the recovery of (\ref{1.12/1.13}) is evident.

\section{The method of reduction of order}

The method of reduction of order was used very effectively by
Nucci \cite{Nucci 96 a} to obtain the complete symmetry group for
the Kepler Problem, the representation of which is mainly in terms
of nonlocal symmetries, in terms of Lie point symmetries of the
reduced system.  The reader is referred to Nucci {\it et al}~\cite{Nucci 00 a} 
for a detailed description of the method.
Essentially the differential equations are written as a system of
first-order ordinary differential equations and the order of the
system is reduced by the removal of ignorable variables.
Eventually  one (or more) variable(s) is (are) eliminated to give
at least one equation of the second order so that Lie's algorithm
for the determination of Lie point symmetries can be implemented
as a finite procedure.  The symmetries obtained are then
translated back to symmetries of the original system.  The
identification of the complete symmetry group is facilitated by
the theorems on transitivity recently obtained by Andriopoulos
{\it et al}~\cite{Andriopoulos 05 a}.

The Ermakov System as presented in (\ref{1.12/1.13}) 
is not satisfactory for the application of the method of reduction 
of order since there is no preferred new variable after the time, 
as an ignorable coordinate, is eliminated.  However, conversion of 
the system to a representation in terms of plane polar coordinates 
through $x = r\cos\theta $ and $y = r\sin\theta $ and recombination 
into radial and tangental components, {\it videlicet}
\begin{subequations}\label{4.1/4.2}
\begin{gather}
 \ddot {r} -r\dot {\theta} ^ 2-r ^ {- 3}\left (\tan\theta +\cot\theta\right) ^ 2 = 0, \label {4.1}\\
 r\ddot {\theta} + 2\dot {r}\dot {\theta} + \half r ^ {- 3}\left
(\tan\theta +\cot\theta\right)' = 0, \label {4.2}
\end{gather}
\end{subequations}
where the prime denotes differentiation with respect to $\theta$,
does provide a good basis for the reduction since ultimately one
can envisage a polar representation of the orbit.

We define the variables $w_1 =r $, $w_2 =\theta $, 
$w_3 = \dot {r} $ and $w_4 = \dot {\theta} $ so that the system  
(\ref{4.1/4.2}) 
can be written in terms of the four first-order ordinary differential equations
\begin{subequations}\label{4.3-4.6}
\begin{gather}
\dot {w}_1  =  w_3, \label {4.3}\\
\dot {w}_2  =  w_4, \label {4.4}\\
\dot {w}_3  =  w_1w_4 ^ 2+ \frac {a (w_2)} {w_1 ^ 3}, \label {4.5}\\
\dot {w}_4  =  - 2\frac {w_3w_4} {w_1}
-\frac {a' (w_2)} {2w_1 ^ 4}, \label {4.6}
\end{gather}
\end{subequations}
where we write $\left (\tan\theta +\cot\theta\right) ^ 2 = a
(\theta) $ to keep a certain compactness of notation.  Obviously
the system is autonomous, i.e.\ $t $ is an ignorable variable.

An obvious candidate as a new independent variable 
is $w_2 $ which we write as $y $.  The ratios of (\ref {4.3}), 
(\ref {4.5}) and (\ref {4.6}) with (\ref {4.4}) give the reduced third-order system
\begin{subequations}\label{4.7-4.9}
\begin{gather}
\displaystyle {\frac {\d w_1} {\d y}}  =  \frac {w_3} {w_4}, \label {4.7}\\
\displaystyle {\frac {\d w_3} {\d y}}  =  w_1w_4+ \frac {a (y)} { w_1 ^ 3w_4}, \label {4.8}\\
\displaystyle {\frac {\d w_4} {\d y}}  =  - 2\frac
{w_3} {w_1}- \frac {a' (y)} { 2w_1 ^ 4w_4}.
\label {4.9}
\end{gather}
\end{subequations}
The next step is to introduce a second-order differential 
equation into the system so that Lie's algorithm becomes finite.  
The obvious variable for elimination is $w_3 $ for from (\ref {4.7}) 
we have the simple relationship $w_3 = w_4w_1'$, where now the prime refers to $y $.

The system of second-order differential equation and first-order differential equation is
\begin{subequations}\label{4.10/4.11}
\begin{gather}
w_4w_1''+w_4'w_1' =  w_1w_4+ \frac {a (y)} { w_1 ^ 3w_4}, \label {4.10}\\
w_4' =  - 2\displaystyle {\frac {w_4w_1'} {w_1}}- \frac
{a' (y)} { 2w_1 ^ 4w_4}.  \label {4.11}
\end{gather}
\end{subequations}
At this stage one can apply Lie's algorithm, 
but some \ae sthetic reshaping of system (\ref{4.10/4.11}) 
is informative.  Equation (\ref{4.11}) can be written in the explicitly conservative form
\begin{gather}
u_2'= 0, \label{4.12}
\end{gather}
where
\begin{gather*}
u_2: = w_1 ^ 4w_4 ^ 2 + a (y). 
\end{gather*}
We recognise $u_2 (y) $ as related to the Ermakov Invariant
appropriate to system (\ref{1.12/1.13}), but we must
recall, as the explicit presence of the independent variable
reminds us, that this is in terms of the independent variable $y $
and not the original independent variable~$t $.  The distinction
is technical since $\d u_2/\d t = u_2'\dot {y} $.

We introduce the new dependent variable into (\ref{4.10}) and
after some manipulation arrive at
\begin{gather}
\left (u_2- a\right)u_1' - \half a'u_1' + u_2u_1 = 0 \label {4.14}
\end{gather}
where we have introduced the new dependent variable $u_1 = 1/w_1 $.

We have reduced the original fourth-order system to a third-order
system comprising one second-order equation~(\ref{4.14}) and
one first-order equation~(\ref{4.12}).  The first-order equation
has the nature of a conservation law.  We recall that the same
type of reduction has been demonstrated for the Kepler Problem,
problems based upon it and the MIC-Z system.  The question which
naturally arises is one of the equivalence of these somewhat
disparate systems.  The resolution of the question is found in
consideration of (\ref{4.14}).  As a linear equation this can be
reduced to that of the simple harmonic oscillator -- the form of
the second-order equation taken in the case of the reduction of
the Kepler Problem -- by means of a Kummer--Liouville
transformation.  The transformation of (\ref {4.14}) to a
nonautonomous oscillator requires simply a change in the dependent
variable.  However, to go from the nonautonomous oscillator to the
simple harmonic oscillator requires a redefinition of the
independent variable.  In the usual area of application, i.e.\
the treatment of the time-dependent oscillator, the introduction
of `new time' is simply a rescaling of a variable without bound.
Here the independent variable has a geometrical meaning and a
rescaling of the polar angle changes the geometry.  Consequently
we must conclude that there is an essential difference between the
two classes of problem despite the formal equivalence obtained
under the procedure of the method of reduction of order.

\section{Symmetries of the original system \\
and its complete symmetry group}

The Lie point symmetries of the system (\ref{4.14}), (\ref{4.12}) are given in composite form as
\begin{gather}
\Gamma = (\alpha +\beta u_1)\partial_y+ (\sigma u_1 ^ 2+\gamma
u_1+\delta)\partial_{u_1} +\epsilon\partial_{u_2}, \label {5.1}
\end{gather}
where
\begin{subequations}\label{5.2}
\begin{gather}
\sigma  =  \frac{\partial\beta}{\partial y} +\frac {a'} {2 (u_2- a)}\beta, \label{5.2a} \\
\gamma  =  \gamma_0+ \frac 12
\left(\frac{\partial\alpha}{\partial y} +
\frac{\alpha a' - \epsilon}{u_2 - a}\right) \label{5.2b},
\end{gather}
\end{subequations}
$\gamma_0 $ is a constant, $\epsilon$ is an arbitrary function of
$u_2$ and $\alpha $, $\beta $ and $\delta $ solutions of the
equations
\begin{subequations}\label{5.3-5.5}
\begin{gather}
\frac{\partial\alpha}{\partial y} +\left [\frac {4u_2} {u_2- a} +
\frac{\partial}{\partial y}\left (\frac {a'} {u_2- a}\right)' -\frac{1}{4} 
\left  (\frac {a'} {u_2- a}\right) ^2\right]\frac{\partial\alpha}{\partial y} \nonumber\\
\qquad {} + \frac 12 \frac{\partial}{\partial y}
\left [\frac {4u_2} {u_2- a} +\frac{\partial}{\partial y}\left (\frac {a'} {u_2- a}\right)
 -\frac {1}{4}\left  (\frac {a'} {u_2- a}\right) ^2\right]'\alpha \nonumber\\
\qquad{}  =  \frac{1}{2(u_2-a)}\left\{\frac{\partial}{\partial y}\left(\frac{a'}{u_2-a}\right)
+\frac{4a}{u_2-a}\right\}, \label {5.3} \\
\frac{\partial\beta}{\partial y} +\frac 12 \frac {a'} {u_2- a}\frac{\partial\beta}{\partial y}
 +\frac 12 \left [\frac {2u_2} {u_2- a} +\left (\frac {a'} {u_2- a}\right)'\right]\beta  =  0,  \label {5.4} \\
\frac{\partial\delta}{\partial y}
-\frac 12 \frac {a'} {u_2- a}\frac{\partial\delta}{\partial y} +\frac {u_2}
{u_2- a}\delta  =  0. \label {5.5}
\end{gather}
\end{subequations}
We note that the equation for $\beta $ is the adjoint of the equation for $\delta $.

Two of the relevant symmetries are \cite{Andriopoulos 02 a} the two
solution symmetries of (\ref {4.14}) given by the $\delta $
equation.  We write these as
\begin{gather}
\Sigma_1  =  s_1\partial_{u_1}, \qquad
\Sigma_2  =  s_2\partial_{u_1}, \label {5.6}
\end{gather}
where $s_1 $ and $s_2 $ are two linearly independent solutions of the $\delta $ equation.

{\samepage We firstly examine the effects of these two symmetries. The general
pair of second-order equation and first-order equation is
\begin{gather*}
u_1''  =  f (y,u_1,u_2,u_1'), \nonumber \\
u_2'  =  g (y,u_1,u_2,u_1'). 
\end{gather*}
The application of $\Sigma_1 ^ {[2]} $ gives
\begin{gather*}
s_1''  = \frac {\partial f} {\partial u_1} + s_1'\frac {\partial f} {\partial u_1'}, \nonumber \\
0  =  s_1 \frac {\partial g} {\partial u_1} +
s_1' \frac {\partial g} {\partial u_1'} 
\end{gather*}
from the solutions of which the two equations in (\ref{5.6}) now
become
\begin{gather}
u_1''  =  \frac {s_1''} {s_1} + F (y,s_1 u_1'-s_1'u_1',u_2),\nonumber \\
u_2'  =  G (y,s_1 u_1' -s_1 u_1',u_2), \label {5.9}
\end{gather}
where $F $ and $G $ are arbitrary functions of the indicated arguments.}

After the application of $\Sigma_2 ^ {[2]} $ to (\ref{5.9}) we have
\begin{gather}
u_1''  =  \frac {s_1''} {s_1}u_1+\frac {1} {s_1}\frac {s_1s_2'' -s_1''s_2} {s_1s_2' -s_1's_2}
\left (s_1u_1' -s_1'u_1\right) + {\cal F} (y,u_2), \nonumber \\
u_2'  =  {\cal G} (y,u_2). \label {5.10}
\end{gather}
 Since
\begin{gather*}
s_1''  =  \frac 12 \frac{a'}{u_2-a}s_1' - \frac{u_2}{u_2-a}s_1, \\
s_2''  =  \frac 12 \frac{a'}{u_2-a}s_2' -
\frac{u_2}{u_2-a}s_2,
\end{gather*}
the system (\ref {5.10}) becomes
\begin{gather*}
(u_2-a)u_1''-\half a'u_1 + u_2u_1 = {\cal F}(y,u_2), \nonumber \\
u_2' =  {\cal G}(y,u_2). 
\end{gather*}

The two remaining symmetries required to specify the system come
from the particular solution of (\ref{5.3}) and its consequence in
(\ref{5.2b}).  The particular solution of (\ref{5.3}) follows from
that of the original equation~(\ref{4.14}) and the use of
Laplace's method of variation of parameters.  The symmetries can
be written as
\begin{gather*}
\Sigma_3 = A_0\partial_y+C_0u_1\partial_{u_1}+\partial_{u_2},
\nonumber \\
\Sigma_4 =
A_0u_2\partial_y+(C_0u_2+\gamma_0)u_1\partial_{u_1}+u_2\partial_{u_2} =u_2\Sigma_3+\gamma_0u_1\partial_{u_2},
\end{gather*} 
where $\gamma_0$ is a constant, $C_0$
is given by (\ref{5.2b}) and
\begin{gather*}
A_0 =
\frac{1}{4W}\left\{\sigma_1^2\int\sigma_2^2\left[\frac{a''}{(u_2-a)^2}+\frac{a'{}^2}{(u_2-a)^3}
+\frac{4a}{(u_2-a)^2}\right]\d
y \right.\nonumber \\
\phantom{A_0 =}{}-\left.\sigma_1\sigma_2\int\sigma_1\sigma_2\left[\frac{a''}{(u_2-a)^2}
+\frac{a'{}^2}{(u_2-a)^3}+\frac{4a}{(u_2-a)^2}\right]\d
y \right. \nonumber \\
\phantom{A_0 =}{}
+\left.\sigma_2^2\int\sigma_1^2\left[\frac{a''}{(u_2-a)^2}+\frac{a'{}^2}{(u_2-a)^3}+\frac{4a}{(u_2-a)^2}\right]\d
y \right\}. 
\end{gather*}
In fact the complexity of the expression for the symmetry is not
of importance.  Rather it is the structure found in $\Sigma_3$ and
$\Sigma_4$.

A symmetry of the original system (\ref{1.12/1.13}) has the general form
\begin{gather*}
\Gamma =\tau\partial_t+\xi\partial_x+\eta\partial_y 
\end{gather*}
in which the variable dependence in the coefficient functions $\tau $, $\xi $ and $\eta $ 
is not specified since for the symmetries of interest to us this is determined 
through the connection to the point symmetries listed in (\ref{5.1}). 
We follow the line of transformations through conversion to polar, i.e.\ $x = r\cos\theta $, $y = r\sin\theta $ and
\begin{gather*}
\Gamma \longrightarrow \tau \partial_t+
\left (\xi\cos\theta +\eta\sin\theta\right)\partial_r+\frac {1} {r}
\left (-\xi\sin\theta +\eta\cos\theta\right)\partial_{\theta} 
\end{gather*}
which for compactness of notation we write as
\begin{gather*}
\Gamma \longrightarrow \tau\partial_t+R\partial_r+\Theta\partial_{\theta}, 
\end{gather*}
expression as the system of four first-order differential equations, i.e.\ $r = w_1 $, 
$\theta = w_2 $, $\dot {r} = w_3 $, $\dot {\theta} = w_4 $ and
\begin{gather*}
\Gamma \longrightarrow \tau\partial_t+R\partial_{w_1} 
+\Theta\partial_{w_2} +\big (\dot {R} -\dot {r}\dot {\tau}\big)\partial_{w_3}
+\big (\dot {\Theta} -\dot {\theta}\dot {\tau}\big)\partial_{w_4}  
\end{gather*}
the reduction of order which means the elimination of $t $ 
and the use of $w_2 = y $ as the independent variable giving the operator
\begin{gather*}
\Gamma \longrightarrow R\partial_{w_1} +\Theta\partial_{y} +
\big(\dot {R} -\dot {r}\dot {\tau}\big)\partial_{w_3}
+\big(\dot {\Theta} -\dot {\theta}\dot {\tau}\big)\partial_{w_4}.   
\end{gather*}

Finally we have the elimination of $w_3$ and the introduction 
of the two new dependent variables $u_1 = 1/w_1 $ and $u_2 = w_1^4w_4^2 - a(y)$ so that
\begin{gather*}
\Gamma \longrightarrow  \big(\Theta -\big(\dot {\Theta} -\dot {\theta}\dot {\tau}\big) a'\big)\partial_{y}  
- u_1^2R\partial_{u_1}  +\big(4w_1 ^ 3w_4 ^ 2R+ 2w_1 ^ 4w_4\big(\dot {\Theta} -\dot {\theta}\dot {\tau}\big)\big)
\partial_{u_2}. 
\end{gather*}

We are now in a position to express the symmetries (\ref{5.6}) 
in terms of the original variables of our toy Ermakov system.  
After a modest amount of manipulation we obtain the symmetries
\begin{gather}
\Delta_1  = \left (\int s_1 \left (\theta\right)r\d t\right)\partial_t
+s_1 (\theta)\left (x\partial_x+y \partial_y\right), \nonumber \\
\Delta_2  = \left (\int s_2 \left (\theta\right)r\d t\right)\partial_t
+s_2 (\theta)\left (x\partial_x+y\partial_y\right), \nonumber\\
\Delta_3  =  \left\{\int\left [\frac {\dot {\Theta}} 
{\dot {\theta}} -\frac {1} {2r ^ 4\dot {\theta} ^ 2} - 2C_0\right]\d t\right\}\partial_t
-\left [C_0x+y\left (A_0+\frac {\dot {a}} {2\left (x\dot {y} -\dot {x}y \right) ^ 2}
 + 2C_0 \dot {a}\right)\right]\partial_x \nonumber\\
\phantom{\Delta_3  =}{}+\left [-C_0y+x\left (A_0+\frac {\dot {a}} {2\left (x\dot {y} -\dot {x}y\right) ^ 2}
 + 2C_0 \dot {a} \right)\right]\partial_y,\nonumber  \\
\Delta_4  =  \left (\left (x\dot {y} -\dot {x}y\right) ^ 2+ a\right)\Delta_3+\frac {\gamma} {x ^ 2+y ^ 2}
\left (x\partial_x+y\partial_y\right), \nonumber\\
\Delta_5  =  \partial_t, \label {5.17}
\end{gather}
where the fifth symmetry is introduced as the symmetry responsible
for the reduction of order by the change of independent variable
from $t $ to $\theta $ and $s_1 $ and $s_2 $ are linearly
independent solutions of (\ref{4.14}).

In (\ref{5.17}) we have a representation of the complete symmetry
group of our original system~(\ref{1.12/1.13}).  The
algebra of the symmetries may be written as $A_1\oplus\left
\{A_2\oplus_s 2 A_1\right\} $ in the notation of the Mubarakzyanov
classification scheme \cite{Morozov 58 a,Mubarakzyanov 63 a, Mubarakzyanov 63
b, Mubarakzyanov 63 c}.  The first Abelian subalgebra 
comprises~$\Delta_5 $ and the second abelian subalgebra the solution
symmetries $\Delta_1 $and $\Delta_2 $.

Note that the four symmetries coming from the system (\ref {4.12}), (\ref {4.14}) 
are nonlocal only in the coefficient of $\partial_t $.  
The symmetries $\Delta_1 $ and $\Delta_2 $ are point in the spatial 
variables as is the case with the nonlocal symmetries used by Krause 
to specify the Kepler Problem.  The symmetries~$\Delta_3 $ and $\Delta_4 $ 
are generalised in the spatial variables.  Given that the original system is autonomous, 
the concentration of the nonlocality in the coefficient of $\partial_t $ is 
not surprising since the integral does not enter of itself into the action of the symmetry on the system.

\section{Conclusion}

In this paper we have considered a very simple example of the
class of problems known as Ermakov systems.  The simplicity of the
choice was deliberate for we were then able to concentrate upon
the important concepts rather than becoming enmired in distracting
manipulations.  Even still the conversion of the symmetries from
their simple representations for the system~(\ref{4.12}) and~(\ref{4.14}) 
to those given in~(\ref{5.17}) was not a completely
trivial exercise.  We have seen that a~representation of the
complete symmetry group of this toy Ermakov system cannot be
achieved by means of Lie point symmetries.  It is necessary to
look to nonlocal symmetries.  Nevertheless our toy system belongs
to the class of superintegrable problems.  It is also of some
interest that by means of the method of reduction of order we were
able to reduce the Ermakov system, fourth-order and nonlinear, to
a third-order system of linear equations.  This same reduction has
been observed for other problems, but there seem to be some
essential differences due to the geometric implications of the
transformation of the independent variable.

As a final remark we recall that the coupled system (\ref{1.12/1.13}) is 
superintegrable.  Other superintegrable systems include 
the simple harmonic oscillator and the Kepler Problem.  The 
former enjoys a richness of symmetry which is not lessened by the 
adjoining of a further dimension or two.  The latter never has a 
sufficient number of Lie point symmetries for its complete specification 
be the dimension two or twenty-two.  Yet it is the classic example of a natural 
superintegrable system.  The Ermakov system considered here occupies a 
middle position in that it has sufficient Lie point symmetries as a 
scalar system to be more than superintegrable and yet does not have 
the requisite number of point symmetries to be completely specifiable.  
Here we have detailed a strategy to reconcile the superintegrablity 
of the system and its ease of complete specification.  The strategies 
for the resolution of particular problems may vary, but they have 
the same underlying basis which is to seek that representation in 
which the superintegrability is supported by an equally generous 
supply of Lie point symmetries for they are the class 
of symmetry with which to work is the easiest.  

The conventional representations of systems such as the cojoined 
Ermakov system considered here are not necessarily those for which 
the representation of the Lie algebraic properties are the simplest.

\subsection*{Acknowledgements}

PGLL thanks the University of KwaZulu-Natal for its continuing
support.

 \LastPageEnding

\end{document}